\documentclass[pre,aps,showpacs,twocolumn]{revtex4}
\usepackage{graphicx,amsmath,bm} 
\begin{document}

\title{Population Dynamics in Spatially Heterogeneous Systems with Drift:
the generalized contact process.}
\author{Jaewook Joo}
\altaffiliation[Present address: ]{Department of Physics,
Pennsylvania State University, University
Park, Pennsylvania, 16803}
\affiliation{Department of Physics, Rutgers University, New Brunswick,
New Jersey, 08854}
\author{Joel L. Lebowitz}
\affiliation{Department of Mathematics and Physics, Rutgers University,
New Brunswick, New Jersey 08854}
\date{\today}
\begin{abstract}
We investigate the time evolution and stationary states of a
stochastic, spatially discrete, population model (contact process)
with spatial heterogeneity and imposed drift~(wind) in one- and
two-dimensions. We consider in particular a situation in which
space is divided into two regions: an oasis and a desert (low and
high death rates).  Carrying out computer simulations we find that
the population in the (quasi) stationary state will be zero,
localized, or delocalized, depending on the values of the drift
and other parameters. The phase diagram is similar to that
obtained by Nelson and coworkers from a deterministic, spatially
continuous model of a bacterial population undergoing convection
in a heterogeneous medium.
\end{abstract}
\pacs{05.50.+q, 05.70.Ln, 87.23.-n}
\maketitle

\section{\label{intro-DCP}Introduction}

In a series of works David Nelson and collaborators
~\cite{nelson:1998,nelson:2000} investigated
the extinction versus survival of bacterial colonies undergoing
convection in a medium with heterogeneous distribution of nutrients.
Space is divided into an``oasis'', with high nutrients,
and a ``desert" with low nutrients. The above model is similar
to models used in mathematical epidemiology in which a population is
divided into subpopulations~(patches) and disease-causing factors are
non-uniformly distributed across patches~\cite{diekmann:2000}.

Nelson $et$ $al$ solved the Kolmogorov-Fisher equation (KFE) with
a convective term, using its similarity to a nonlinear
Schr$\ddot{o}$dinger equation with a non-hermitian
perturbation~\cite{murray:1980,nelson:1998,nelson:2000}. The 
eigenvalue spectrum of this equation has a diverging localization
length with a sharp mobility edge. This mobility edge is
identified with the Fisher velocity, the speed of propagation in
the KFE. This means that while at high convection velocity all
bacteria are transported into the desert and become extinct, at
low velocity the surviving population is localized near the oasis.
If the desert can support a non-zero steady state population
extinction is replaced by a delocalized steady state.

It is tempting to consider the continuum deterministic equations
of Nelson $et$ $al$~\cite{nelson:1998,nelson:2000} as a coarse
grained description of a more microscopic stochastic model of this
system
~\cite{marro:1999,kampen:1997,durrett:1994a,lebowitz:1985-86a,
lebowitz:1988,durrett:1994,liggett:1985,grimmett:1989,
shnerb:2000}. A potential candidate for such a model is a
generalized contact process~(CP) on a lattice which includes both
drift and spatial heterogeneity. Using Monte Carl simulations, we
find that this stochastic model does in fact have a phase diagram
which is qualitatively similar to that obtained from Nelson's
deterministic continuum approach. Various transitions between
phases (extinct, localized and delocalized) are identified.

Stochastic modeling of the discrete microscopic space-time events
underlying the type of physical phenomena considered in
~\cite{nelson:1998,nelson:2000,diekmann:2000} raises many
conceptual and methodological questions.  These are further
complicated by the practical limitations of computer
simulations---the size of the system, the time it can be followed
and the number of samples are all strictly limited and are
generally small compared to the macroscopic space-time scales of
the physical phenomena. Experience nevertheless indicates that
judicious microscopic modeling, which requires choosing system
sizes and simulation times large enough to well represent
``typical'' behavior plus ``normal'' fluctuations, can add to our
understanding of the macroscopic
phenomena~\cite{marro:1999,kampen:1997,durrett:1994a,liggett:1985}.
On the other hand the simulation time must not be too large so as
to avoid rare events such as extinction of populations, which will
always occur eventually in finite systems.  These, however, occur
on a time scale that grows exponentially with the size of the
system and so will almost never be seen at intermediate time
scales. We do not attempt here any deep analysis of these
important issues but refer to the literature
\cite{marro:1999,durrett:1994,liggett:1985}.

The more modest goal of this paper is to investigate the effect of
locomotion with drift, in an inhomogeneous spatial environment, on
the behavior of the stochastic contact process. This is a
frequently used model for population dynamics, spread of
epidemics, etc.
\cite{diekmann:2000,murray:1980,durrett:1994a,liggett:1985,
grimmett:1989,marro:1999}.
We consider in particular an environment consisting of a single
oasis surrounded by a patch of desert several times the size of
the oasis. By using periodic boundary conditions, this is
equivalent to considering a periodic array of such patches which,
quite unrealistically, are microscopically identical. We first
describe the basic contact process with additional spatial jumps
and find that these jumps have the effect of destroying
correlations. This brings the system closer to its mean field
behavior. We then investigate the oasis-desert system in one and
two dimensions. Hydrodynamical scaling which connects this
stochastic model to continuum deterministic models is discussed
briefly.

\section{\label{model-DCP}Description of Model}

%[the basic contact process]
The contact process~(CP) is frequently used to model population
dynamics where the population lives on the $d$-dimensional cubic
lattice
$Z^{d}$~\cite{liggett:1985,marro:1999,durrett:1994a,durrett:1994,
grimmett:1989}.
At each site $x$ of the lattice there is an organism which can
exist in one of two states $\{0,1\}$: we will refer to $0$ as
dead or $1$ as alive. Let $\eta_{x} = 0,1$ be the variable which
specifies the state at a site $x$ of a $d$-dimensional simple
cubic lattice ${Z}^d$ and let $\underline{\eta}=\{ \eta_{y}\}$, $y
\in Z^{d}$, be the configuration of the system at some given time.
The stochastic dynamics of the process are specified by the
following rates:
\begin{eqnarray}
\eta_{x}=1 \rightarrow \eta_{x}=0 &\text{ at rate }& \alpha(x),
\label{transitionrule1}
\\
\eta_{x}=0 \rightarrow \eta_{x}=1 &\text{ at rate }&
(\lambda/2d)\sum_{y:|y-x|=1} \eta_{y}
\label{transitionrule2}
\end{eqnarray}
where $2d$ is the number of neighbors of site $x$. Equivalently, a
live site dies at rate $\alpha(x)$ or attempts to give birth to a
new bacterium on a randomly chosen neighboring site at rate
$\lambda/2d$. It succeeds if that site is empty, otherwise nothing
happens. We are assuming here for simplicity that the birth rate
is spatially uniform, although of course one might expect both to
depend on local conditions. In particular one expects higher birth
rates in the oasis than in the desert. What is most important,
however, is just the ratio of birth rate to death rate, hence the
effect we are modeling should be essentially captured by just
considering variations in the latter.

As there must be at least one live site to propagate, the all-zero
state is absorbing, i.e., if $\eta_{x}=0$ for all $x$ at any time
$t$ it will stay like that forever
~\cite{liggett:1985,grimmett:1989,marro:1999}. When
$\alpha(x)=\alpha$ is constant, the evolution depends only on the
ratio $\lambda/\alpha$. For large $\lambda/\alpha$ a live
population can survive. As $\lambda/\alpha$ decreases there is a
transition from survival to extinction at a critical value of this
ratio which depends on the dimension $d$ of the lattice. In units
of $\alpha=1$, $\lambda_{c} \simeq 3.30$ in one-dimension and
$\lambda_{c} \simeq 1.65$ on the square
lattice~\cite{marro:1999,liggett:1985}.

In the mean field approximation, corresponding to the neglect of
correlations between different sites, one can write a deterministic
equation for the density of live sites.  Thus, for the spatially
homogeneous case, with density $\langle \eta_x(t) \rangle = \rho(t)$,
we have
\begin{equation}
\frac {d \rho(t)}{dt} = -\alpha \rho + \lambda(1 - \rho) \rho^2.
\label{3}
\end{equation}
Starting with some non zero density $\rho(0) \leq 1$ the long time
solution of Eq.~(\ref{3}) is  $\rho(\infty) = 0$ if
$\alpha/\lambda>1$, and $\rho(\infty) = 1 - \alpha/\lambda$, if
$\alpha/\lambda<1$, and so $\lambda^{(MF)}_c = 1$.  This shows in
particular that $\lambda_c^{(MF)} < \lambda_c$ for $d=1,2$ (or
indeed for any finite $d$). This can be understood by noting that
correlations between live (occupied) sites are positive for the
contact process in the stationary state, i.e.: $\langle \eta_x
\eta_y \rangle \geq \langle \eta_x \rangle \langle \eta_y
\rangle$. Thus correlations inhibit growth, since live sites give
birth only on unoccupied neighboring sites.

We now add to the CP a mechanism for the movement of the bacteria
via a Kawasaki exchange kinetics
~\cite{liggett:1985,grimmett:1989,marro:1999,durrett:1994a,lebowitz:1985-86a,lebowitz:1988,durrett:1994}.
This permits a bacterium at site $x$ to move (jump) to a
neighboring site $y$, $|x-y|=1$. The exchanges conserve the total
number of individuals in the system. These two processes can be
considered as follows. A clock at each live site $x$ rings at a
rate 1, i.e., the probability that it will ring in a very small
time interval $dt$ is equal to $dt$. When it rings the bacterium
at that site attempts to jump to a neighboring site with
probability $P_{K}$, dies with probability $(1-P_{K})\alpha(x)/c$,
or attempts to give birth to a bacterium on a randomly chosen
neighboring site with probability $(1-P_{K}) \lambda /(2d c)$, and
with probability $(1 - P_K)[1 - (\alpha(x) + \lambda/2d)/c]$
nothing happens.  The constant $c$ is chosen as small as possible
to make all probabilities non-negative, i.e. $c = \lambda/2d +
\max_x \alpha(x)$.  We note further that if the bacteria decides
to jump or give birth on a neighboring site which is already
occupied, nothing happens.

If a jump is decided upon, the appropriate neighboring site is chosen
as follows: in one dimension a bacterium at site $x$ attempts to jump
to the site $x+1$ with probability $q$ and to the site $x-1$
with probability $1-q$, $0 \leq q \leq 1$.
In two dimensions the symmetry of the Kawasaki
exchanges along the $y$ axis is broken while the exchanges along the $x$ axis
are symmetric. A bacterium at site $(x,y)$ attempts to jump to site $(x,y+1)$ with
probability $q/2$ and to site $(x,y-1)$ with probability $(1-q)/2$
while it tries to jump to site $(x+1,y)$ and to site $(x-1,y)$
with equal probability $1/4$.
When $q > 1/2$ there is a drift to the right or up (left or down for $q<1/2$).
In all cases jumps succeed if the target site is empty, otherwise
nothing happens.

When $q=1/2$ particles jump in each direction with equal
probability. This corresponds to a symmetric stirring. For
constant $\alpha$ and $q=1/2$ it has been shown in
Ref.~\cite{lebowitz:1985-86a} that in the  limit, $P_{K}
\rightarrow 1$ one can rescale space and time appropriately to
obtain rigorously a continuum reaction-diffusion equation for the
density.  A space time-scaling, leading to a reaction-diffusion
equation with drift holds also for $q > \frac{1}{2}$; see Section
VI. This means that when the jumps become very frequent they
destroy the correlations between sites.  Subsequently the critical
value of $\lambda$ assumes its mean-field value in this limit.

In general the configuration probability distribution,
$\mu(\underline{\eta},t)$, evolves in time according to the master
equation,
\begin{equation}
\frac{\partial \mu(\underline{\eta},t)}{\partial t}=\biggl[
P_{K}L_{K}+(1-P_{K}) L_{G}\biggr]\mu(\underline{\eta},t)
\label{generator-master}
\end{equation}
where $L_{G}$ describes the Glauber dynamics, changing $\eta_{x}$ to $1-\eta_{x}$,
while $L_{K}$ describes the Kawasaki exchanges between neighboring sites $|x-y|=1$
~\cite{marro:1999,lebowitz:1985-86a}.
%One has formally that
%\begin{eqnarray}
%L_{G}&=&\sum_{x}\biggl(G_{x}-1 \biggr)c(\underline{\eta},x),
%G_{x}\mu(\underline{\eta})=\mu(\underline{\eta^{x}})
%\\
%L_{K}&=&\sum_{|x-y|=1}\biggl(H_{xy}-1\biggr)c(\underline{\eta};x,y),
%H_{xy}\mu(\underline{\eta})=\mu(\underline{\eta^{xy}})
%\end{eqnarray}
%where $\mu( \underline{\eta^{xy}} )$ is the configuration probability
%distribution obtained from $\underline{\eta}$ after the exchange.

\section{One Dimensional System}

\subsection{Homogeneous System}

\begin{figure}
\begin{center}
\includegraphics[height=6cm,angle=0]
{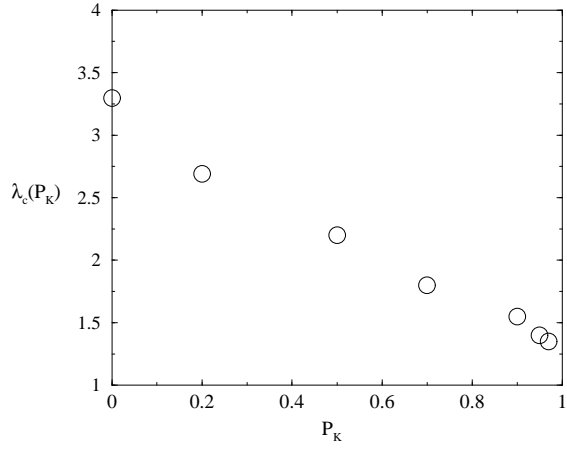}
%{Fig.042004-1.eps}
\caption{\label{fig1}The critical values $\lambda_{c}(P_{K})$ as
a function of the jump probability $P_{K}$ for $q=0.5$.
Results are extrapolated to $N=\infty$ (open circles) from system sizes of $N=50, 100, 200, 300, 500$.
The average is taken over $10^{3}$-$10^{4}$ different realizations starting with random initial
configurations at density.}
\end{center}
\end{figure}

%[symmetric exchange]
We carried out Monte Carlo simulations of the contact process with
a homogeneous death rate, $\alpha=1$, and with symmetric Kawasaki
exchanges for various jump probabilities $P_{K}$. The system was a
ring containing 50, 100, 200, 300, 500 sites (with periodic
boundary conditions). Results were obtained by taking averages
over $10^{3}$-$10^{4}$ realizations, each starting with an initial
configurations obtained by putting a bacterium at each site with
probability $\rho_{0}$. For each realization $\rho_{0}$ was chosen
uniformly in the interval (0,1). As is well known the only steady
state in the finite system is the absorbing state corresponding to
$\eta_{x}=0$ for all $x$~\cite{liggett:1985,marro:1999}. It is
only in an infinite system where there can be steady states with a
nonvanishing population density. It is therefore necessary to
extrapolate the critical values of the infinite system from the
results of the simulations of finite systems~\cite{marro:1999}.
The extrapolated critical values of the CP in one dimension are
shown in Fig.~\ref{fig1} as a function of $P_{K}$. The simulations
indicate an approximately linear decrease of $\lambda_{c}$ with
$P_{K}$ and are consistent with the expectations that stirring
reduces correlations.  As already noted  in the limit of rapid
stirring the system behaves in a mean field way with
$\lambda_{c}(P_{K})\rightarrow 1$ as $P_{K} \rightarrow 1$
~\cite{lebowitz:1985-86a}.

\begin{figure}
\begin{center}
\includegraphics[height=6cm,angle=0]
{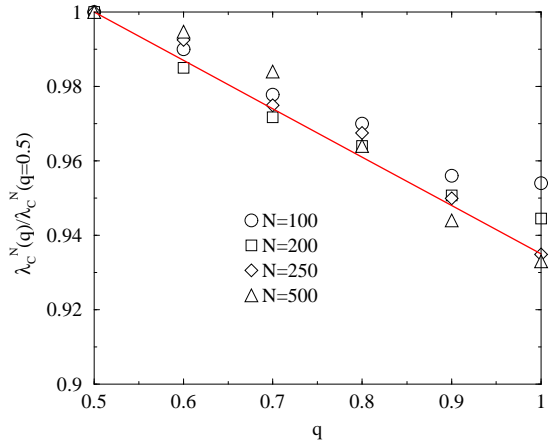}
%{Fig.112404-1.eps}
\caption{\label{fig2}The critical value of $\lambda^{N}_{C}(q)$
for a homogeneous system as a function of drift $q$ in one dimension.
$P_{K}=0.9$ and system sizes of $N=100,200,250,500$ are used.
The function that is used for the solid line is $1-\gamma(q-0.5)$ with $\gamma=0.13$.
}
\end{center}
\end{figure}

We also simulated the homogeneous contact process with drift
in one dimension. The critical values $\lambda_{c}$, for $P_{K}=0.9$
are plotted in Fig.~\ref{fig2} for different values of $q$.
We find that the critical value $\lambda_{c}$ decreases
approximately linearly as $q$ increases to 1;
here again the drift helps survival
by reducing correlations between occupied sites.  It does this by
randomly exposing new empty sites on which births can take place. Such
an effect does not occur for a deterministic ballistic motion which
satisfies Gallilean invariance.
%{\it How about $\lambda_{c}$ at fixed $q>1/2$, as $P_{K} \rightarrow 1$?}
%{\it Apparently in two dimensional phase diagram in Fig.8,
%$\lambda_{c}$ does not depend on $q$. investigate it in the case of
%homogeneous death rate}

\subsection{Spatial Heterogeneity}

We introduce two types of spatially varying death rates in one dimension:
\begin{eqnarray}
\alpha_{I}(x)=&&1+\epsilon \sin (\frac{2\pi x}{N})
\label{deathrate1}
\\
\alpha_{II}(x)=&& \left \{
  \begin{array}{cc}
  1-\epsilon  \! \text{, $x$ $\in$ Oasis } \! \\
  1+\epsilon  \! \text{, $x$ $\in$ Desert }\!
  \end{array} \right.
\label{deathrate2}
\end{eqnarray}
where $x \in \{1,2,...,N\}$ with periodic boundary condition,
$0<\epsilon < 1$.
The oasis is located in $\{x|1\leq x \leq N/5\}$ and
the desert is everywhere else.
For $\epsilon=0$, $\alpha=1$ sets the unit of time.

\begin{figure}
\begin{center}
\includegraphics[height=6cm,angle=0]
{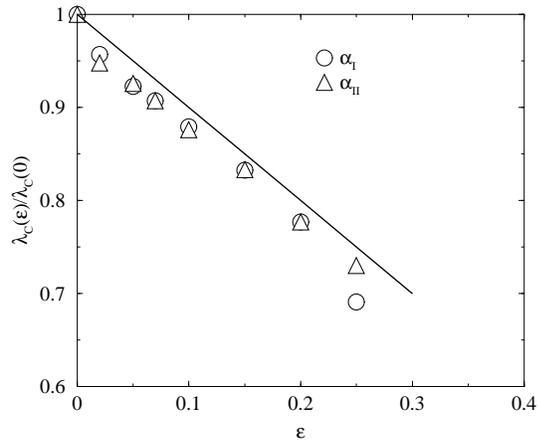}
%{Fig.042104-1.eps}
\caption{\label{fig3}The ratio of the critical values in heterogeneous system
to those in the homogeneous system,
$\lambda_{c}(\epsilon)/\lambda_{c}(0)$.
The parameters used for the simulation are $P_{K}=0$
and the results are extrapolated to $N=\infty$ (open symbols) from
$N=50,100,200,500,1000$. The average is taken over $10^{3}-10^{4}$
different realizations. The function that is used for the solid line
is $1-\epsilon$.
}
\end{center}
\end{figure}

For both types of spatial variation we used simulations to obtain the
critical values $\lambda_{c}(\epsilon)$, defined as the smallest value
of $\lambda$ for which a population can survive as a function of
$\epsilon$.
In the absence of exchanges, $P_{K}=0$,
the critical value $\lambda_{c}(\epsilon)$ decreases
monotonically
as $\epsilon$ increases. The ratios of $\lambda_{c}(\epsilon)/\lambda_{c}(0)$
are plotted against $\epsilon$ in Fig.~\ref{fig3}.
We find
\begin{equation}
\label{ratio}
\lambda_{c}(\epsilon)/\lambda_{c}(0)\simeq 1-\epsilon=\textrm{Min}_{x}\alpha(x).
\end{equation}

\begin{figure}
\begin{center}
\includegraphics[height=6cm,angle=0]
{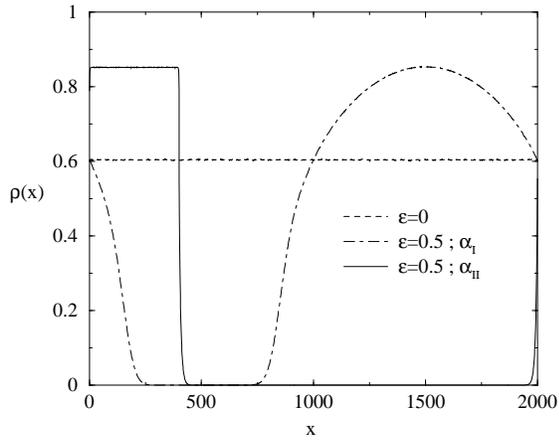}
%{Fig.080204-4.eps}
\caption{\label{fig4}The stationary density $\bar{\rho}$(x)
on an one dimensional ring of size $N=2000$ for $P_{K}=0$ and
$\lambda=4$. $\alpha_{I}(x)$ and $\alpha_{II}(x)$ with $\epsilon=0.5$
are used, respectively.}
\end{center}
\end{figure}

Eq.~(\ref{ratio}) suggests that the system behaves, for large $N$, as
consisting of approximately independent local regions (patches) with different values
of the death rate $\alpha(x)$.
If that were true exactly we would have
$\lambda_{c}(\epsilon)/\lambda_{c}(0)=1-\epsilon$.
The stationary population density $\bar{\rho}(x)$ is
plotted in Fig.~\ref{fig4} for both heterogeneous and
homogeneous death rates using a value of $\lambda$ which
is smaller than $\lambda_{c}$ for $\alpha=1+\epsilon$ but larger than
$\lambda_{c}$ for $\alpha=1-\epsilon$.

\subsection{Spatial Heterogeneity with Exchanges}

In this section we investigate the steady states of the contact process
with both Kawasaki exchange kinetics and a spatial heterogeneity
of the form $\alpha_{II}(x)$ given in Eq.~(\ref{deathrate2}).
The nature of the stationary states of the system now depend on the strength of
the drift $q$.

In the simulations we chose $P_{K}=0.9$ so
that we could see the change of the stationary states as a response
to the variation of the drift, while keeping the total simulation time,
which should increase~\cite{lebowitz:1985-86a} proportionally to $1/(1-P_{K})$ manageable.
The critical values $\lambda_{c}(q)$ were obtained for
different values of the drift $q$ in systems of size,
$N=50,100,200,500$ and then were extrapolated to those in an infinite system.

\begin{figure}
\begin{center}
\includegraphics[height=6cm,angle=0]{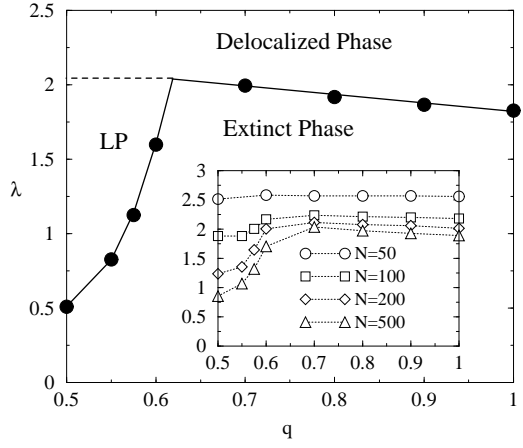}
%{Fig.112304-1.eps}
\caption{\label{fig5}The $\lambda$-$q$ phase diagram of the contact process with
drift and spatial heterogeneity, $\alpha_{II}(x)$ with $\epsilon=0.5$, in one dimension.
Here $P_{K}=0.9$.
The critical curve (solid line with filled circles), obtained by
extrapolation to the infinite
system from $N=50,100,200,500$(shown in the inset),
separates the survival (delocalized or localized)
from extinct phase. LP stands for the localized phase.}
\end{center}
\end{figure}

The $\lambda$-$q$ phase diagram is shown in Fig.~\ref{fig5} for $\epsilon=0.5$.
The critical curve (solid line), $\lambda_{c}(q)$ separates
the survival and extinction regions. For $\lambda \leq \lambda_{c}(q)$
the extinct phase, $\eta_{x}=0$ for all $x$, is the only steady
state.
For $\lambda>\lambda_{c}(q)$ there is a steady state with a finite population
density which will be approached as $t \rightarrow \infty$,
starting with a non-zero initial density.
There are two survival steady states: a localized and a delocalized one.
In the localized survival phase the population density is non-zero only in the oasis and its vicinity while in the
delocalized survival phase the population density is non-zero everywhere.
As $\epsilon$ decreases the region of the localized phase in the phase diagram shrinks.
In the limit $\epsilon=0$ the critical curve collapses onto the slanted critical
curve of the contact process with drift and homogeneous death rate
$\alpha=1$ [see Fig.~\ref{fig2}].

\begin{figure}
\begin{center}
\includegraphics[height=6cm,angle=0]{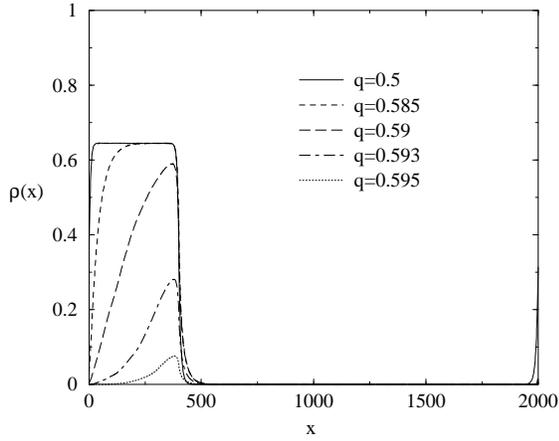}
%{Fig.073004-15.eps}
\caption{\label{fig6}Drift-driven transition from the localized phase to the extinct
phase. The parameters used for the simulation are : $N=2000$,
$\lambda=1.5$,
$\alpha_{II}(x)$ with $\epsilon=0.5$ and $P_{K}=0.9$. Drift varies from $q=0.5$ where population is localized
in the vicinity of the oasis to $q=0.598$
where the population becomes extinct.}
\end{center}
\end{figure}

\begin{figure}
\begin{center}
\includegraphics[height=6cm,angle=0]{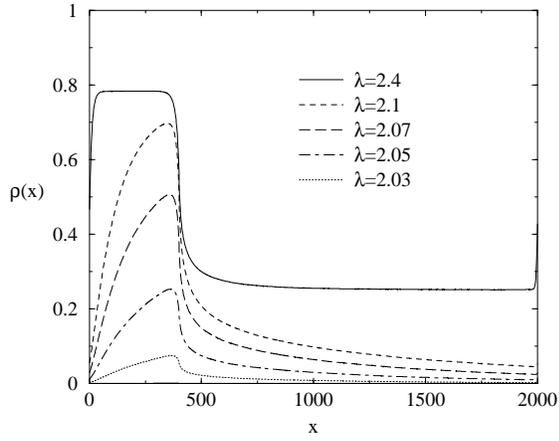}
%{Fig.073004-12.eps}
\caption{\label{fig7}Transition from the delocalized survival phase to the extinct phase.
The parameters used for the simulation are system size of $N=2000$,
$q=0.9$,
$\alpha_{II}(x)$ with $\epsilon=0.5$ and $P_{K}=0.9$. Birth rate varies from $\lambda=2.4$ where population is
extended everywhere to $\lambda=1.9$ where the population becomes
extinct.}
\end{center}
\end{figure}

The dependence of the stationary density profile $\rho(x)$
on $q$ and $\lambda$ are
shown in Figs.~\ref{fig6} and \ref{fig7}.
Fig.~\ref{fig6} shows how the population in the oasis is decreased
as the wind (drift $q$) increases at $\lambda=1.5$.
For this value of $\lambda$ the oasis is able to sustain a population
while the desert is not: $\lambda_{c2}(\epsilon)< \lambda <\lambda_{c1}(\epsilon)$ where
$\lambda_{c1}(\epsilon)$ [or $\lambda_{c2}(\epsilon)$] is defined as the critical value for
the CP with symmetric Kawasaki exchange with $P_{K}=0.9$ and homogeneous death rate
$\alpha=1+\epsilon$ [or $\alpha=1-\epsilon$].
There is a critical drift, $q_{c}(\lambda)$ for each fixed $\lambda$, above which the population becomes
extinct in the long time limit.
When the drift is weak, [$q<q_{c}(\lambda)$] the surviving population is localized near the oasis.
When the drift is strong, [$q>q_{c}(\lambda)$] the population is "blown'' into the desert
and becomes extinct.
We see from Fig.~\ref{fig5} (where $\epsilon=0.5$), that $\lambda_{c1}(\epsilon=0.5)\simeq 2$,
$\lambda_{c2}(\epsilon=0.5)\simeq 0.5$ and
$q_{c}(\lambda_{c1})\simeq 0.6(1)$.
%For Fig.~\ref{fig5}, where $\epsilon=0.2$, $\lambda_{c1}\simeq 1.75$,
%$\lambda_{c2} \simeq 1.25$ and $q_{c}(\lambda_{c1})=0.5(5)$.

Fig.~\ref{fig7} shows the density profile for a strong drift, $q=0.9$, at
different $\lambda$'s.
The condition for survival, when the drift is stronger than the critical
drift, is that
the desert should be able to sustain a population, i.e.,
$\lambda>\lambda_{c1}(\epsilon)$.
For $\lambda>\lambda_{c1}(\epsilon)$ the population survives everywhere
and its steady state is the delocalized phase.
The condition for this survival is very much similar to that of
the CP with homogeneous death rate of
$\alpha=1+\epsilon$ (desert) and with strong drift($q=0.9$).

When $q<q_{c}(\lambda)$, a transition from the localized to the delocalized survival
phase can be achieved by making $\lambda$ cross $\lambda_{c1}$,
the dashed line in $\lambda-q$ phase diagram in Fig.~\ref{fig5}.
When $\lambda$ is very close to $\lambda_{c1}$, it is also possible
to have a transition from the localized phase
(through the extinction phase) to the delocalized phase for constant $\lambda$.
This is due to the slanted critical curve in the $\lambda-q$ plane:
One can draw a horizontal line crossing three phases in the $\lambda-q$ plane:
the population is localized at small drift, (undergoes an extinction transition,)
 and becomes delocalized at strong drift.

\section{Spatial Heterogeneity with Exchanges in Two-dimensions}

In $d=2$ the exchange kinetics impose a drift along the $y$
direction. We consider a periodic box of length $N$ and choose the
region of oasis to be $\{(x,y)|2N/5<x \leq 3N/5, 2N/5<y \leq
3N/5\}$ and the desert everywhere else. We considered square
lattices of sizes $N^{2}=30^{2},50^{2},70^{2}$,$100^{2}$, with
periodic boundary conditions, for $\epsilon=0.5$ in Eq.
~(\ref{deathrate2}) and $P_{K}=0.9$, and extrapolated the critical
curve to that in an infinite system.

Note that the steady states in two dimensions also depends on the
shape of oasis. An oasis of narrow width
along the direction of the drift will yield a very different steady states
from one having a narrow width perpendicular to the direction of the drift.
One can expect that in the former case the steady states will be similar to
those of the CP with heterogeneous death rate $\alpha_{II}(x)$ with symmetric exchanges
in one dimension while for the latter case the steady states will be
similar to those of the CP with $\alpha_{II}(x)$
under the drift in one dimension.

\begin{figure}
\begin{center}
\includegraphics[height=6cm,angle=0]{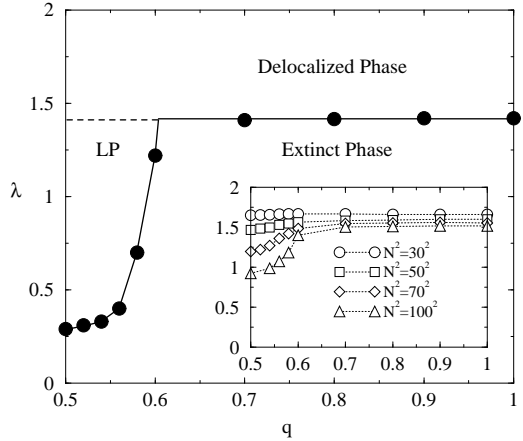}
%{Fig.112304-2.eps}
\caption{\label{fig8}The $\lambda$-$q$ phase diagram of the contact process
with drift and spatial heterogeneity in two dimension.
The parameters used for the simulation are $P_{K}=0.9$, $\alpha_{II}(x)$ with $\epsilon=0.5$.
The critical curve (solid line with filled circles) is obtained by
extrapolation to the infinite system from
$N^{2}=30^{2},50^{2},70^{2},100^{2}$(shown in the inset).
The oasis is a square of length $N/5$. LP stands for the localized phase.
}
\end{center}
\end{figure}

\begin{figure}
\begin{center}
\includegraphics[height=6cm,angle=0]{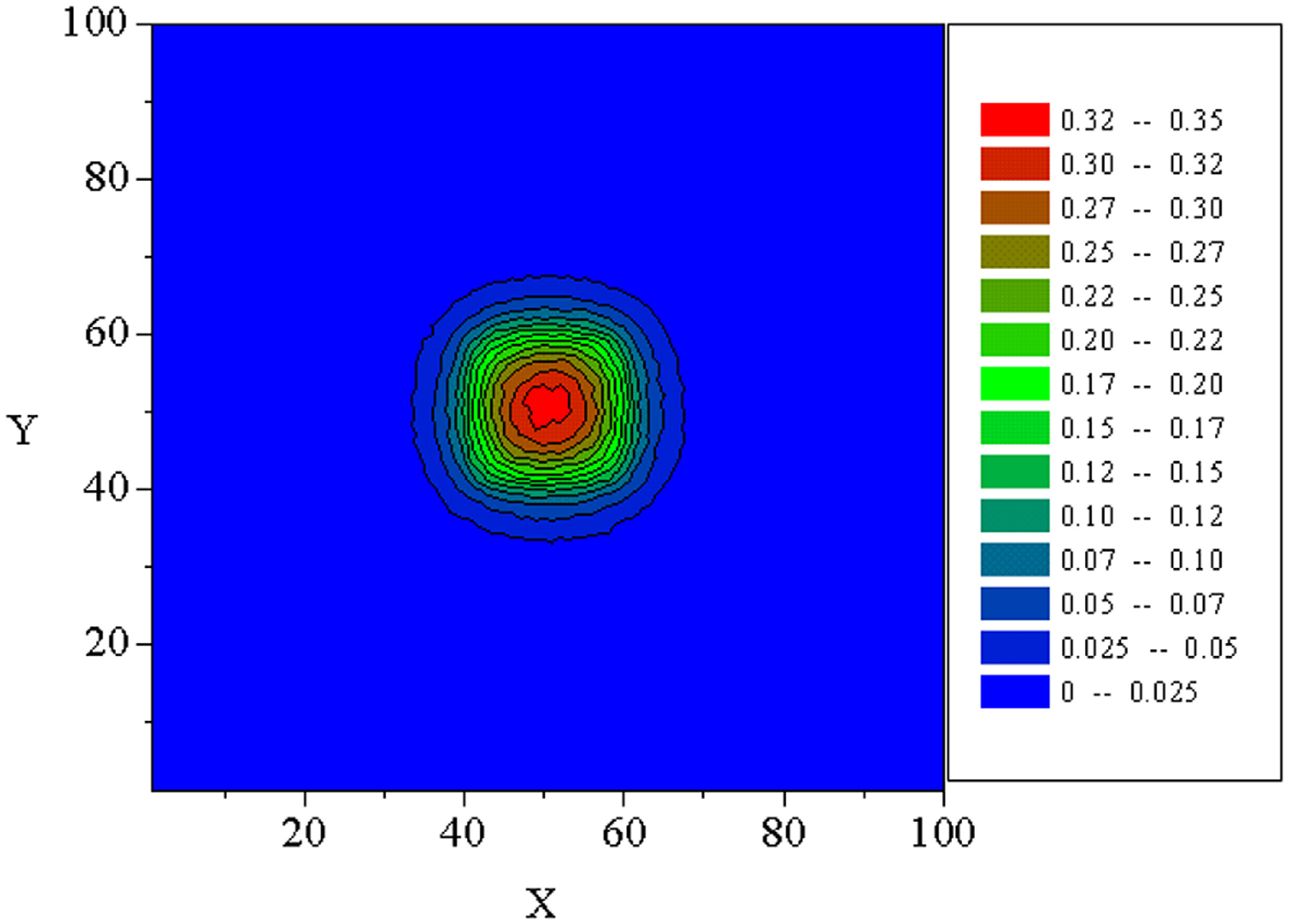}
%{L1Q0.5N100.eps}
\caption{\label{fig9}The localized phase in two dimension
 for $\lambda=1.05$, $q=0.5$, $\alpha_{II}(x)$ with $\epsilon=0.5$, $P_{K}=0.9$ and $N^{2}=100^{2}$.
}
\end{center}
\end{figure}

\begin{figure}
\begin{center}
\includegraphics[height=6cm,angle=0]{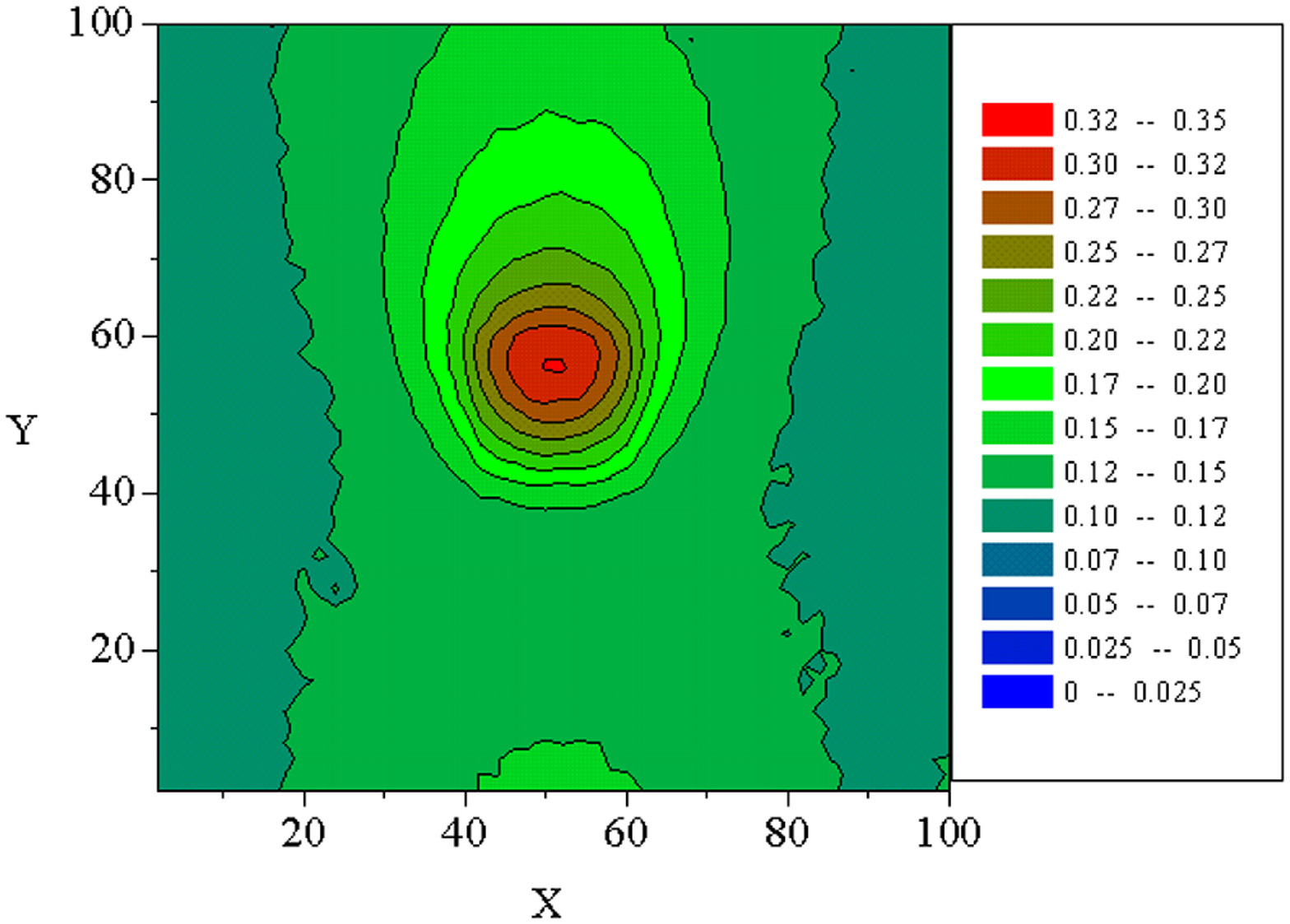}
%{L1.75Q0.7N50.eps}
\caption{\label{fig10}The delocalized phase in two dimension
 for $\lambda=1.75$, $q=0.7$, $\alpha_{II}(x)$ with $\epsilon=0.5$, $P_{K}=0.9$ and $N^{2}=100^{2}$.
}
\end{center}
\end{figure}

The $\lambda$-$q$ phase diagram for the CP with drift and with spatial heterogeneity
in two dimensions is plotted in Fig.~\ref{fig8} for various system sizes
with the oasis a square with sides of length $N/5$.
The localized [delocalized] phase in two dimensions is identified
by the presence of a large cluster of occupied sites around the oasis, which is ``not large
enough'' [``large enough''] to reach both boundaries of the system along the $x$ or $y$ direction.
The delocalized, or localized survival phase and the extinction phase are separated
by the critical
curve $\lambda_{c}(q)$. The shape of that critical curve is similar to that in
one dimension.  This is not too surprising since, unlike equilibrium
systems, nonequilibrium systems like the CP, can have phase
transitions also in one dimension with similar phase diagrams
~\cite{marro:1999}.   We can therefore expect similar phase transitions between three phases
by varying $(\lambda, q)$: examples of localized and delocalized phases are
depicted in Figs.~\ref{fig9} and \ref{fig10}.

\section{Time-evolution of the Population Density $\rho_{t}(x)$}

To study the time evolution of the population density $\rho_{t}(x)$
undergoing drift in a heterogeneous system,
we performed Monte Carlo simulations in one dimension with $N=2000$, $P_{K}=0.9$
and $\epsilon=0.5$ in Eq.~(\ref{deathrate2}).
We selected a set of $(\lambda,q)$ parameters from different
steady states in the $\lambda$-$q$ phase diagram in Fig.~\ref{fig5} and evolved the system
from various initial conditions.

In the localized and in the delocalized phases
we started with a population density $\rho_{0}=0.01$ randomly placed.
For each set of parameters we simulated $10^{4}$ realizations of the process with
the above initial conditions.
After 100, 200, 500, 2000 Monte Carlo time-steps we measured the population density $\rho_{t}(x)$.
(In one Monte Carlo time step each site is visited once for updating its state, on average.)

\begin{figure}
\begin{center}
\includegraphics[height=6cm,angle=0]{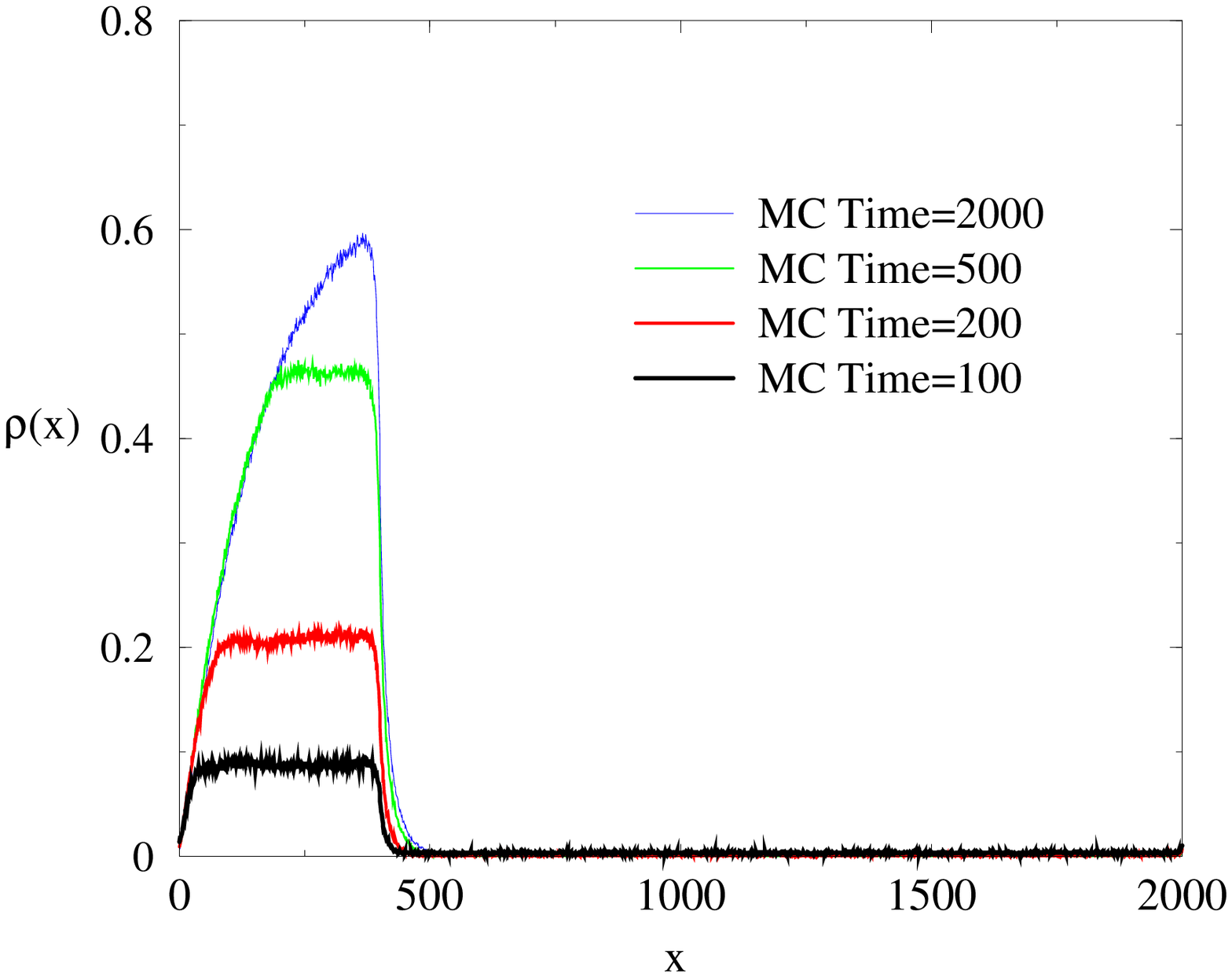}
%{Fig.112304-3.eps}
\caption{\label{fig11}Time evolution of population density $\rho_{t}(x)$ for
$\lambda=1.5$, $q=0.59$, $\alpha_{II}(x)$ with $\epsilon=0.5$, $P_{K}=0.9$ and $N=2000$,
i.e., for a weak drift in the localized phase.
$10^{4}$ realizations starting with 1 percent of randomly distributed population are
averaged at given Monte Carlo time-steps.
}
\end{center}
\end{figure}

\begin{figure}
\begin{center}
\includegraphics[height=6cm,angle=0]{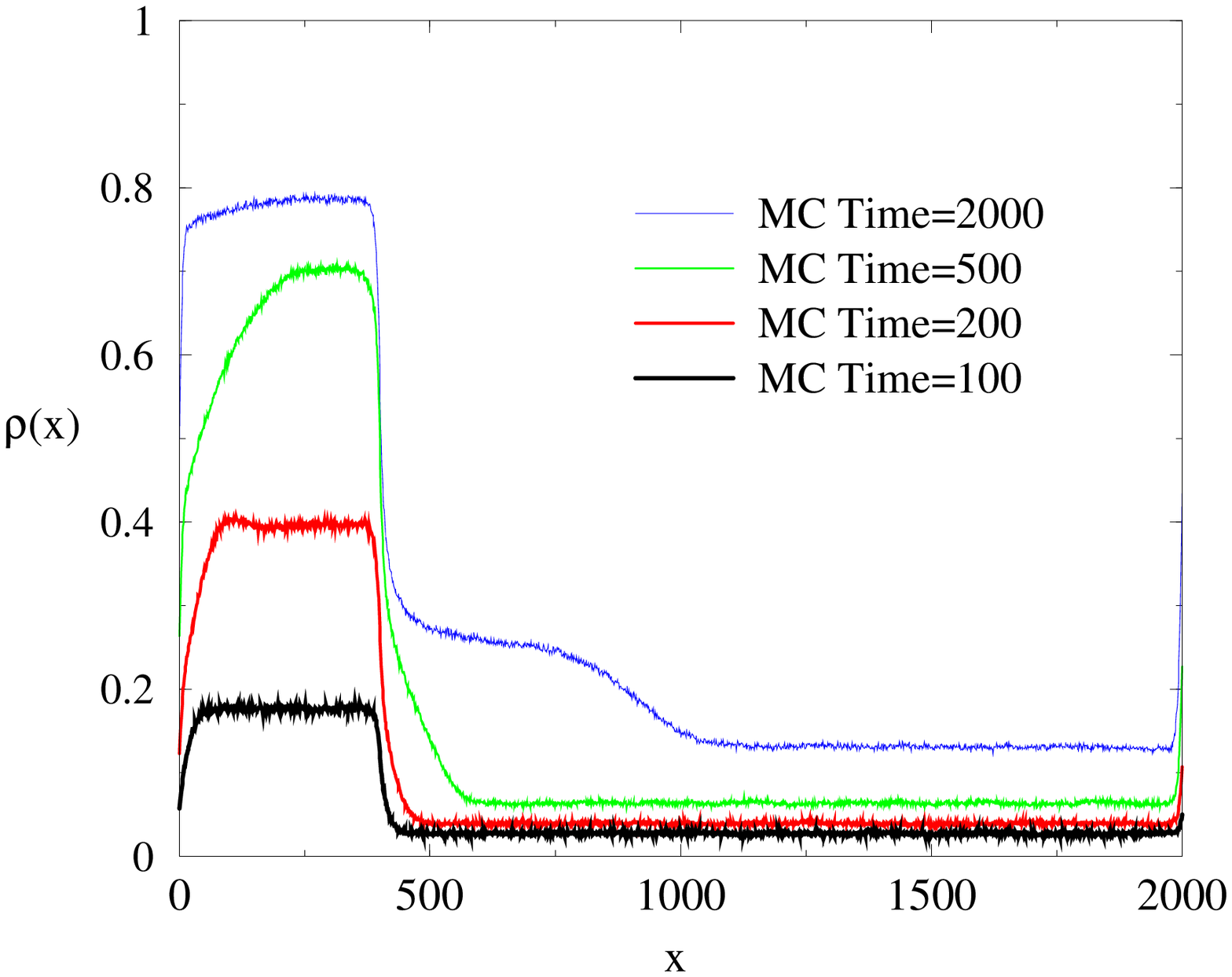}
%{Fig.112304-4.eps}
\caption{\label{fig12}Time evolution of population density $\rho_{t}(x)$ for
$\lambda=2.5$, $q=0.59$, $\alpha_{II}(x)$ with $\epsilon=0.5$, $P_{K}=0.9$ and $N=2000$,
i.e., for a weak drift in the delocalized phase.
$10^{4}$ realizations starting with 1 percent of randomly distributed population
are averaged at given Monte Carlo time-steps.}
\end{center}
\end{figure}

Figs.~\ref{fig11} and~\ref{fig12} show the time-evolution of a
population density under a weak drift ($q=0.59$) for a small birth
rate ($\lambda=1.5$) and for a large birth rate ($\lambda=2.5$).
For the small birth rate ($\lambda=1.5$) the population grows
mostly in the vicinity of the oasis and its density profile is
skewed to the right due to the presence of the drift: under a
stronger drift the population will die out in this case. For the
large birth rate ($\lambda=2.5$) the population density profile at
2000 Monte Carlo time-steps decreases gradually in the desert as
one goes further away from the oasis. Note that in the strong
drift case, $q=0.9$ shown in Fig.~\ref{fig7} the population
density profile is flatter in the desert for $\lambda=2.4$.

\section{Comparison of the Contact Process and the Kolmogorov-Fisher Equation}

De Masi $et$ $al$~\cite{lebowitz:1985-86a} derived a reaction diffusion equation
for the symmetric exclusion process with added Glauber dynamics
by rescaling time and space, $x=\epsilon j a$ ($j=...,-1,0,1,...$)
and $t=\epsilon^{2} \tau$, $a$ and $\tau$ being the microscopic
lattice spacing and $\tau$ being the time in microscopic units (jump time)
and then taking the $\epsilon \rightarrow 0$ limit with $(1-P_{K}) \sim \epsilon^{2}$.

We expect that under suitable scalings the model considered in this paper leads,
when $P_{K} \rightarrow 1$, to
a continuum macroscopic equation of the form~\cite{lebowitz:1985-86a,bahadoran:2004}
\begin{eqnarray}
\label{MFlebowitz}
\frac{\partial P_{t}(x)}{\partial t} &=& D\nabla^{2}P_{t}(x)-v
\nabla P_{t}(x)(1-P_{t}(x)) \nonumber\\&+&
(\lambda-\alpha(x))P_{t}(x)-\lambda P_{t}^{2}(x)
\end{eqnarray}
where $P_{t}(x)$ is the bacteria density at time $t$ at position
$x$, $D$ is a diffusion constant and $v$ is a convection velocity
where $D$ and $v$ can be obtained from the microscopic model.
(Note that on the time and space scales considered here there is
no contribution to $D$ from the birth and death proven, i.e.\ $D
\to 0$ when $P_K \to 0$.)

Eq.~(\ref{MFlebowitz}) resembles the continuum modified KFE used by
Nelson\cite{nelson:1998,nelson:2000},
which generalizes the KFE to account for convection and an inhomogeneous growth rate,
\begin{equation}
\label{MFnelson}
\frac{\partial c(x,t)}{\partial t} =D_{N} \nabla^{2} c(x,t)
-v_{N} \nabla c(x,t)+U(x)c(x,t)-b c^{2}(x,t),
\end{equation}
where $c(x,t)$ is the bacteria population density, $D_{N}$ is the
(Nelson) diffusion constant of the bacteria, $U(x)$ is the spatially
varying growth potential corresponding to $(\lambda-\alpha(x))$ in Eq.~(\ref{MFlebowitz}),
$v_{N}$ is an externally imposed (Nelson) convection, and $b$ is a parameter that limits the
population number density to a maximum saturation value.
The only difference between Eq.~(\ref{MFlebowitz}) and Eq.~(\ref{MFnelson})
is the nonlinear convection term in Eq.~(\ref{MFlebowitz}).
Because our interest lies mainly in the location of the critical curve,
we can linearize Eq.~(\ref{MFlebowitz}), while keeping a nonlinear
reaction term.
Then the coefficients in the two equations can be matched with each other.
This makes it possible to compare the simulation results with those
from the differential equation. We find good qualitative agreement between two approaches,
even though we don't present the results of comparison here.

\section{Conclusions}
We studied the steady states and dynamical behavior of the contact
process with asymmetric Kawasaki exchange kinetics and with
spatially varying death rate. This is a stochastic, spatially
discrete, version of the deterministic continuum Kolmogorov-Fisher
equation with convection velocity, investigated by Nelson and
collaborators~\cite{nelson:1998,nelson:2000}. The phase diagrams
from both the stochastic and the deterministic cases, birth rate
$\lambda$ vs $q$, are similar: two (localized and delocalized)
survival phases are separated from the extinction phase by a
critical curve, $\lambda_{c}(q)$.  As is true in many other
nonequilibrium situations, the difference between one and two
dimension is small. Simulations of the time-evolution of the
population under drift in one dimensional heterogeneous systems
show how a population evolves to the localized or the delocalized
steady states, depending on the birth rate.

%After elementary calculations and simple comparisons we can express
%the Fisher velocity $v_{F}$ in terms of microscopic parameters:
%$v_{F}=2\sqrt{(\lambda_{c}-1+\epsilon)/\lambda_{c}}$
%$=v\sqrt{P_{K}/D\lambda_{c}(1-P_{K})}$.
%One can obtain the Fisher velocity from the typical KFE, a
%linearized Eq.~(\ref{MFlebowitz}) without convection term [see the Ref.~\cite{murray:1980}]
%and set the Fisher velocity, $v_{F}=2\sqrt{(\lambda_{c}-1+\epsilon)/\lambda_{c}}$,
%equal to the prefactor of $\nabla P_{t}(x)$, $v\sqrt{P_{K}/D\lambda_{c}(1-P_{K})}$.
%If $q>q_{c}(\lambda)$, $dP_{t}(x)/dt < 0$ leading the population to
%extinction.
%Otherwise the population survives.
%Then the critical curve takes the form $\lambda_{c}(q)
%\sim A(q-1/2)^{2}+B$, which is approximately the shape of the critical curve in Fig.~\ref{fig6}
%and \ref{fig11}.
%
%
%Both $\lambda_{c1}$ and $\lambda_{c2}$ can be expressed in terms of microscopic parameters.
%The sign and magnitude of mobility edge, defined as the eigenvalue of the fastest growing delocalized state
%[see Ref.~\cite{nelson:1998,nelson:1999}],
%depends on average $\langle (\lambda-\alpha(x))/\lambda \rangle$ and its sign
%changes at the boundary
%between localized and delocalized state, i.e., at $\lambda_{c1}$.
%Thus $\lambda_{c1}=\langle \alpha_{II}(x) \rangle$.
%For $\epsilon=0.5$ $\lambda_{c1}= 1.3$ while from simulations
%$\lambda_{c1}=2$. The definition of $\lambda_{c2}$ in both cases
%is the critical value for the system with homogeneous death rate $1-\epsilon$ with stirring probability
%$P_{K}=0.9$.

\begin{acknowledgments}
Both JJ and JJL thank Pablo Ferrari for helpful comments. 
JJ thanks DIMACS for support and acknowledges the support of grants
NSF DBI 99-82983 and NSF EIA 02-05116. JJL was supported by NSF
DMR-01-279-26 and by AFOSR AF 49620-01-1-0154.
\end{acknowledgments}

\end{document}